\documentstyle [12pt] {article}
\title{A VARIANT AND EXPLANATION OF THE DELAYED CHOICE EXPERIMENT }
\author{Vladan Pankovi\'c, Milan Predojevi\'c \\
Department of Physics, Faculty of Sciences \\21000 Novi Sad, Trg Dositeja Obradovi\'ca 4, Serbia \\
vdpan@neobee.net}

\date {}
\begin{document}
\maketitle

\vspace {0.5cm}


\vspace{1cm}

\begin {abstract}

In this work we suggest a variant of the remarkable Wheeler's
delayed choice gedanken experiment. In our experiment, single
photon described by a superposition state with two terms
dynamically interacts with an atom. Preparation of the atom in any
of two excited states can be realized practically in the last
moment before interaction. For atom in the first excited state
there is practically none dynamical interaction between atom and
photon so that the interference effects on the photon can be
detected later by a photo plate. For atom in the second excited
state dynamical interaction between photon and atom causes
certainly the stimulated emission of a new photon that moves
coherently with the first photon. Both photons do a super-system
described by an entangled quantum state. But in this case photo
plate, that realizes simultaneously sub-systemic measurement at
any photon, does not detect interference effects. We suggest a
simple explanation of given as well as original variant of the
delayed choice experiment in full agreement with standard quantum
mechanical formalism.
\end {abstract}
\vspace {1.5cm}

In this work we shall suggest a variant of the remarkable
Wheeler's delayed choice gedanken experiment [1], [2] recently
realized by Jacques et al. [3]. Also, we shall suggest a simple
explanation of given as well as original variant of the delayed
choice experiment in full agreement with standard quantum
mechanical formalism [4]-[7].

In our experiment, quantum state of the single photon, after
quantum mechanical dynamical interaction between this photon and a
fixed half-silvered mirror, obtain the following form
\begin {equation}
  |p> = 2^{\frac {1}{2}}|R> + 2^{\frac {1}{2}}|T>          .
\end {equation}
It represents a quantum superposition of two, equivalently
probable, superposition terms that, roughly speaking, propagate
along different trajectories. Quantum state $|R>$ describes the
photon reflected by half-silvered mirror while quantum state $|T>$
describes the photon that goes through half-silvered mirror. But
both terms, reflected by corresponding two fixed mirrors, go
simultaneously nearly an atom that holds three energy levels.
Given reflections do not break superposition of the trajectories
so that it can be supposed that before interaction with atom
photon is still described by quantum state (1) (now $|R>$ and
$|T>$ include change of the trajectories after reflection by
corresponding mirrors). Suppose that preparation of the atom in
arbitrary excited quantum state (eigen state of the energy
observable) can be realized very quickly, practically in the last
moment before passing of the superposition terms, i.e., roughly
speaking, photon. (Technical details of given preparation have not
principal role so that they will not be discussed in this work.)

Suppose, firstly, that atom is really prepared in the first
excited quantum state. Suppose, also, that for atom in the first
excited quantum state, there is practically none quantum
mechanical dynamical interaction between atom and photon. So, if
it is chosen that atom be really prepared in the first excited
quantum state then interference between superposition terms of the
photon can be detected later by a photo plate. Precisely, photo
plate, for a series, precisely a statistical ensemble of the
measurements, will detect the interferent statistical distribution
of single photons. All this admits that propagation of the photon
between half-silvered mirror and atom (in the first excited
quantum state), exactly described by quantum state (1), be
interpreted consequently classically as propagation of a classical
wave.

Suppose, secondly, that atom is really prepared in the second
excited quantum state. Suppose, also, that for atom in the second
excited quantum state spontaneous emission of a photon can be
neglected. Further, suppose that for atom in the second excited
quantum state quantum mechanical dynamical interaction between
photon and atom causes certainly stimulated emission of a new
photon that moves coherently with the first photon. Both photons,
"initial", $i$, and "new", $n$, do a quantum super-system exactly
described by an entangled (correlated) quantum state
\begin {equation}
  |i+n> = 2^{\frac {1}{2}}|1>_{i}\otimes |1>_{n} + 2^{\frac {1}{2}}|2>_{i}\otimes |2>_{n}          .
\end {equation}
Here $|1>$ represents the quantum state corresponding to the first
trajectory of a photon after interaction between initial photon
and atom. Also, $|2>$ represents the quantum state corresponding
to the second trajectory of a photon after interaction between
initial photon and atom. Finally, $\otimes$ represents the
tensorial product. In this way first term in (1) describes two
photons, "initial" and "new", that propagate coherently along the
same, first trajectory. Also, second term in (1) describes two
photons, "initial" and "new", that propagate coherently along the
same, second trajectory. Of course, terms "initial" and "final"
have only formal meaning, since it is impossible determinate which
of two photons is initial and which is new, obtained by stimulated
emission.

As it is well-known theoretically [4]-[7] and experimentally [8],
in the entangled quantum state the exact separation of the quantum
super-system in its quantum sub-systems, i.e. photons in our case,
is quantum mechanically unadmitable. In other words there are such
so-called super systemic quantum observable whose measurement can
affirm that given super-system was really before measurement in
the entangled quantum state but not in a mixture of the
non-entangled quantum states. Nevertheless, as it is well-known
too [7], there is a limitation of the analysis of the quantum
super-system in so-called sub-systemic quantum observable by
corresponding sub-systemic measurements. In respect to such
sub-systemic analysis any of the quantum sub-systems was
effectively approximately before measurement in a mixture
(so-called second kind mixture) of the quantum states. It implies
that quantum super-system was effectively approximately before
measurement in a mixture (so-called second kind mixture) of the
non-entangled quantum states. It is very important to be pointed
out that given approximate sub-systemic description of the quantum
super-system and sub-systems is non-numerical. It means that
mentioned mixtures are effectively exact. I.e. they do not
represent pure quantum states with weakly interfering but not
exactly non-interfering terms. In this way there are two
discretely different and complementary levels of the accuracy of
the analysis of a quantum super-system. First one is completely
exact, super-systemic. Second one is incomplete, sub-systemic. By
further analysis it is always possible chose one of given two
levels, either super-systemic or sub-systemic. But it is
impossible that both mentioned levels of the analysis be satisfied
simultaneously. Also, it is impossible that any delayed choice of
the exact, correlated quantum state of the quantum super-system be
realized. Or, it is impossible that super-systemic measurement
changes dynamical evolution of the quantum super-system in the
past even if, of course, given measurement changes discretely, in
the moment of the measurement, quantum state of the super-system.

So, in a sub-systemic analysis appropriate for measurement of the
coordinates of $i$ and $n$, i.e. for a typical detection by a
fixed photo plate, i and n as quantum sub-systems of the quantum
super-system $i+n$ are described respectively by the following
mixtures, precisely by the following statistical operators
\begin {equation}
  \hat {\rho}_{i}= \sum_{j=1}^{2}  \hspace{0.4cm} _{n}<j|i+n><i+n||j>_{n} = \frac {1}{2}|1>_{i \hspace{0.2cm}i}<1| + \frac {1}{2} |2>_{i {\hspace{0.2cm}}i}<2|
\end {equation}
\begin {equation}
  \hat {\rho}_{n} = \sum_{j=1}^{2} \hspace{0.4cm} _{i}<j|i+n><i+n||j>_{i} = \frac {1}{2} |1>_{n \hspace{0.2cm} n}<1| + \frac {1}{2} |2>_{n \hspace{0.2cm}n}<2|             .
\end {equation}
None of given statistical operators corresponds to superposition,
i.e. interference of the quantum states $|1>$ and $|2>$. In the
same sub-systemic analysis super-system $i+n$ is described by the
following mixtures of non-entangled quantum states, precisely by
the following statistical operator
\begin {equation}
   \hat {\rho}_{i+n}=  \frac {1}{2} |1>_{i \hspace{0.2cm} i}<1|\otimes|1>_{n \hspace{0.2cm}n}<1| + \frac {1}{2} |2>_{i \hspace{0.2cm} i}<2|\otimes |2>_{n \hspace{0.2cm} n}<2|   .
\end {equation}

So, it can be supposed that photo plate in a typical detection
procedure interacts simultaneously and independently with $i$ and
$n$. It means that photo plate realizes simultaneously
sub-systemic measurements of the coordinate at $i$ and $n$
statistically distributed by $\hat {\rho}_{i}$(3) and $\hat
{\rho}_{n}$(4). But, of course, it is quantum mechanically really
impossible differ initial and new photon. For this reason photo
plate detects, in fact, non-interferent statistical distribution
of single photon with double intensity.

All this admits that propagation of the photon between
half-silvered mirror and atom (in the second excited quantum
state), exactly described by quantum state (1), be interpreted
consequently classically as the propagation of a classical
particle.

So, we suggested an experiment within which the choice of
arbitrary of two possible different, precisely complementary,
experimental arrangements can be done. Also, we presented exact
quantum mechanical description of any of given experimental
arrangements. It is not hard to see that our experiment represents
a variant, precisely an extension of remarkable Wheeler's delayed
choice gedanken experiment [1], [2]. Namely, in original delayed
choice experiment, before detection, there is no entangled quantum
state. In our experiment, before detection, there is an entangled
quantum state. But, from exact quantum mechanical view point, it
does not represent any principal difference between original and
our form of the delayed choice experiment. More precisely, exact
quantum mechanical formalism admits superposition on a simple
(without sub-systems) quantum system as well as on a quantum
super-system (with sub-systems).

Wheeler's delayed choice experiment [1], [2] represents,
metaphorically speaking, final emphasis of Feynman's dramatic
interpretation [9], [10] of early discussion between Einstein and
Bohr on Young's double-slit experiment [4],[5] and , generally, on
the conceptual foundations of the quantum mechanics. As it is
well-known Einstein and Bohr suggested two completely opposite
conceptions of the quantum mechanics that can be simply called
hidden variable and standard (Copenhagen) respectively.

Simply speaking, hidden variable theories suppose the following.
(We shall not analyze concrete details of given theories but only
such basic principles characteristic for most of given theories.)
Usual space (of the coordinates) represents the basic physical
space. Physical object in this space has corresponding strictly
determined form. (For many physical objects the usual space can be
formally generalized by a phase space.) In different concrete
hidden variables theories this form can be different, e.g. like a
classical particle, classical wave or some combination of the
classical particle and wave. Dynamical state of the physical
object evolves strictly deterministically according to
corresponding non-linear dynamical law. Quantum mechanics yields a
simplified description of the physical object. Hilbert space,
quantum mechanical dynamical state representing unit norm vector
in this space, and, linear, precisely unitary symmetric (that
conserves unit norm and superposition) quantum mechanical dynamics
represent only formal (abstract) mathematical constructions. These
constructions are appropriate for simple theoretical reproduction
of the facts obtained by simple measurements. Collapse, i.e.
superposition or, precisely, unitary symmetry breaking in the
measurement process represents a typical dynamical symmetry
breaking. It can be explained by an initial statistical
distribution of the exact, small non-linear dynamical terms
effectively unobservable, i.e. hidden at the approximate, quantum
mechanical, level of the analysis accuracy. Obviously, hidden
variables theories are conceptually analogous to classical
mechanics or classical field theory.

Principal problem of the hidden variables theories represents
well-known fact [11], [8] that dynamics of a hidden variables
theory consistent with experiments must be super-luminal. It seems
physically implausible. Moreover, according to original
formulation of the Wheeler's delayed choice gedanken experiment
[1], [2] and its recent experimental affirmation by Jacques et al.
[3] it seems that dynamical effects of a hidden variables theory
must change the dynamical state of the quantum system in the past.
It, practically, breaks entirely basic physical concept of the
dynamical evolution. Rejection of the collapse, i.e. attempt of
the reduction of the collapse at a dynamical effect leads toward
rejection of practically any reasonable dynamical concept. In our
variant of the delayed choice experiment except all mentioned
problems there is the following additional problem for hidden
variables theories. Namely, "initial" and "new" photon are
coherent and there is no quantum mechanical dynamical interaction
between given photons. For this reason hidden variables theories
can very hardly explain why photo plate does not detect
interferent statistical distribution of single photon with double
intensity.

Standard quantum mechanical formalism considers that Hilbert space
represents the basic physical space. Physical object in this space
is completely described by quantum mechanical dynamical state that
strictly deterministically evolves according to unitary symmetric
quantum mechanical dynamics. Physical characteristics of the
quantum system are presented by average values of corresponding
Hermitian operators, so-called observables. Unitary symmetry of
the quantum mechanical dynamics expresses that all bases in the
Hilbert space, representing corresponding quantum mechanical
reference systems (referential frames) are equivalent, i.e. that
there is no absolute quantum mechanical referential frame for
description of the unitary symmetric quantum mechanical dynamics.
Nothing more is necessary for exact quantum mechanical description
of the quantum system.

But, of course, there is collapse, i.e. superposition breaking on
the quantum system by measurement, i.e. by interaction between
measured quantum system and measuring apparatus. As it is
well-known [6], collapse cannot be modeled by quantum mechanical
dynamical interaction between measured quantum system and
measuring apparatus. More generally, supposition that collapse
represents an exact quantum phenomenon leads immediately or
intermediately either toward hidden variables theories or toward
metaphysical conceptions (e.g. immaterial Abstract Ego or
consciousness of the human observer [6] etc.) Nevertheless, Bohr
suggested phenomenologically, without concrete formalization, that
collapse represents only a relative and effective phenomena. It
appears (without numerical approximation) on the measured quantum
system only in respect to classically, i.e. not quite exactly
(including corresponding numerical approximations) , quantum
mechanically, described measuring apparatus - generator of the
collapse. Completely exact, unitary quantum mechanical dynamical
interaction between measured quantum system and measuring
apparatus can be simply successfully modeled [6]. This, von
Neumann's dynamical interaction entangles (correlates) measured
quantum system and measuring apparatus, i.e. it extends the
superposition from measured quantum system at the quantum
super-system, measured quantum system + measuring apparatus. Such
dynamical interaction, i.e. corresponding entangled state is
principally different from results of the measurement, i.e.
corresponding mixture of the non-entangled states. Nevertheless,
given dynamical interaction is in full agreement with supposition
on the quantum mechanical dynamical evolution as unique exact way
of the change of the quantum mechanical dynamical state.

In this way there is two simultaneously existing, but principally
and discretely different, i.e. complementary, descriptions of the
interaction between measured quantum system and measuring
apparatus. First one is exact von Neumann's dynamical interaction.
Second one is effective, approximate and corresponds to the
collapse by measurement. Any of these two descriptions can be
chosen, i.e. used at corresponding level of the analysis accuracy.
Such choice expresses remarkable Bohr's principle of the
complementarity, i.e. relative boundary between measured quantum
system and measuring apparatus (metaphorically called principle of
the psycho-physical parallelism).

So, Bohr suggested none concrete formalization of the collapse.
But he supposed implicitly that collapse represents an effective,
local phenomena. Precisely, effective appearance of the collapse
by classical approximation of the quantum mechanics he compared
with effective appearance of Newton's gravitational force in a
local Euclidian approximation of the Riemannian space-time curved
by gravitational field in the general theory of relativity. Bohr
said: "Before concluding I should still like to emphasize the
bearing of the great lesson derived from general relativity theory
upon the question of physical reality in the field of quantum
theory. In fact, notwithstanding all characteristic differences,
the situation we are concerned with in these generalizations of
classical theory presents striking analogies which have often been
noted. Especially, the singular position of measuring instrument
in the account of quantum phenomena, just discussed, appears
closely analogous to the well-known necessity in relativity theory
of upholding an ordinary description of all measuring processes,
including sharp distinction between space and time coordinates,
although very essence of this theory is the establishment of new
physical laws, in comprehension of which we must renounce the
customary separation of space and time ideas. The dependence of
the reference system, in relativity theory, of all readings of
scales and clocks may even be compared with essentially
uncontrollable exchange of the momentum or energy between the
objects of measurement and all instruments defining the space-time
system of the reference, which in quantum theory confront us with
the situation characterized by the notion of complementarity. In
fact this new feature of natural philosophy means a radical
revision of our attitude as regards physical reality, which may be
paralleled with the fundamental modification of all ideas
regarding the absolute character of physical phenomena, brought
about general theory of relativity." [5]

There is a possibility [12], [13] that collapse by quantum
measurement be considered as an especial case of the general
formalism of the spontaneous (non-dynamical) symmetry breaking
[14]-[16]. Namely, by spontaneous symmetry breaking exact dynamics
is always stable and its symmetry is always conserved. But an
approximate dynamics, discretely different from the exact, can be
globally (in whole space) unstable and locally (in some domains of
the space) stable. For this reason, at the approximate level of
the analysis accuracy only, it seems effectively that dynamical
state is statistically localized and that symmetry is broken. In
fact given symmetry is only effectively hidden at the approximate
level of the analysis accuracy. For example such situation exists
in Weinberg-Sallam's theory of the gauge symmetric electro-weak
interaction. Here quantum field dynamics cannot be solved exactly.
Also, here approximate theory represents theory of the small
perturbations that diverges nearly false vacuum and converges
nearly local minimums of the potential energy density. Obviously,
spontaneous (non-dynamical) symmetry breaking (effective hiding)
is principally different from dynamical symmetry breaking.

Concretely, in the quantum mechanics quantum mechanical dynamical
state is always dynamically stable and its unitary symmetry, i.e.
superposition is always exactly conserved. Meanwhile quantum
mechanical dynamical state seems classical mechanically
dynamically stable only in the well-known wave packet
approximation. That represents a numerical approximation. Then
quantum system seems like a classical particle. In an especial
case a quantum mechanical dynamical state representing
superposition of the weakly interfering wave packets is also
always dynamically stable and given superposition is always
exactly conserved. But, from classical mechanical view point, it
is not hard to see that given superposition is globally unstable
and locally (nearly any wave packet center) stable. In this way
conditions for spontaneous superposition breaking (effective
hiding) at the classical mechanical level of the analysis accuracy
become satisfied and spontaneous superposition breaking occurs.
Given spontaneous superposition breaking will be called
self-collapse. It can be observed that self-collapse can occur
only over basis defined by given weakly interfering wave packets.
Any other basis, whose vectors represent non-trivial superposition
of the vectors from the first basis, turns spontaneously, by
spontaneous superposition breaking, in the first basis.

Suppose now that there are two quantum sub-systems that do a
quantum super-system. Suppose, also, that given super-system is
exactly quantum mechanically described by an entangled quantum
mechanical dynamical state. This entangled state is always exactly
quantum mechanically stable and superposition at given
super-system is always exactly conserved. Suppose, further, that
one quantum sub-system can be sub-systemically described by a
second kind mixture of the quantum states representing weakly
interfering wave packets. Then entangled quantum mechanical
dynamical state of the quantum super-system can be analogously
spontaneously broken in the statistical mixture of the
non-entangled quantum states. It, simply speaking, corresponds to
an effective change of the second kind mixture in the first kind
mixture on both quantum sub-systems. Or, it corresponds to
self-collapse on the first quantum sub-system and relative
collapse on the other quantum sub-system. It can be pointed out
the relative collapse on the second quantum sub-system appears
only over basis correlated with basis of the weakly interfering
wave packets of the first quantum sub-system. Suppose, finally,
that correlation between two quantum sub-systems has been realized
by a typical von Neumann's dynamical interaction. This interaction
realizes one-to-one correspondence, i.e. correlation , between one
basis in Hilbert space of the first quantum sub-system and one
basis Hilbert space of the second quantum sub-system. Also, many
different pairs of so-correlated sub-systemic bases can exist.
Nevertheless, there is only one correlation between basis of the
weakly interfering wave packets of the first quantum sub-system
and corresponding basis of the second quantum sub-system. It is
not hard to see that in the described case self-collapsed quantum
sub-system can be considered as the measuring apparatus while
relative-collapsed quantum sub-system can be considered as
measured quantum system. Also, dynamical interaction between given
two quantum sub-systems in appropriate approximation with
self-collapse and relative collapse can be considered as the
measurement. By given measurement only such observable is really
measured if its  eigen basis represents the basis over which
relative collapse on the measured quantum system. It implies that
non-commutative observables cannot be simultaneously measured.
Moreover, it implies that within quantum mechanics an analogy of
the Bell inequality [11] cannot be formulated at all. It means
that quantum mechanics represents consequently a sub-luminal or,
in the relativistic generalization, luminal physical theory.  All
this represents a consequent and complete formalization of Bohr's
(Copenhagen) interpretation of the quantum mechanics.

Experimental affirmation of the Copenhagen interpretation needs
detection of the entangled quantum state of the quantum
super-system, measured quantum system + measuring apparatus. It,
for a macroscopic measuring apparatus, represents, to this day,
technically hardly realizable aim. Nevertheless, as it has been
discussed in [12], given entangled state would be observed in an
experiment suggested by Marshall et al. [17]. In this experiment
quantum superposition on a practically macroscopic mirror in
Michelson's interferometer is considered. Precisely there is a
periodically changeable quantum mechanical dynamical interaction
between one photon that propagates through interferometer and a
movable mirror. During first sub-period given interaction can be
considered as a typical von Neumann's interaction that correlates
two trajectories of the photon and two quantum states of the
movable mirror. One of the quantum state of the mirror represents
a wave packet while other represents a superposition of two weakly
interfering wave packets. Both states are mutually weakly
interfering too. It practically means that here mirror can be
considered as a typical measuring apparatus for detection of the
photon trajectory according to Copenhagen interpretation. Really,
any additional sub-systemic measurement at the photon will point
out that trajectories of the photon are non-interfering. During
second sub-period given interaction between photon and mirror
de-correlates super-system, photon + mirror, in the dynamically
independent sub-systems, photon and mirror. Then additional
measurement on the photon will point out that trajectories of the
photon are interfering. It is possible only under condition that
quantum super-system, photon + mirror, during the first sub-period
has been really described by entangled quantum state but not by a
mixture of non-entangled quantum states.

Original Wheeler's delayed choice experiment can be explained by
standard quantum mechanical formalism, i.e. Copenhagen
interpretation in the following way. After quantum mechanical
dynamical interaction between photon and half-silvered mirror
photon is exactly described by quantum state $|p>$ (1). It
represents a superposition of two quantum states, $|R>$ and $|T>$,
or, roughly speaking, it corresponds to propagation of a classical
wave over two trajectories simultaneously. Given superposition can
be later observed intermediately by a series, i.e. statistical
ensemble of the measurements, precisely photon coordinate
detections by a fixed photo plate. Word "fixed" implies that there
is no momentum exchange between photon and photo plate, precisely
that there is no quantum mechanical dynamical correlation between
photon trajectories and photo plate trajectories.

Suppose, meanwhile, that photo plate is movable. Word "movable"
implies that there is momentum exchange between photon and photo
plate, precisely that there is quantum mechanical dynamical
correlation between photon trajectories and photo plate
trajectories. It means that super-system, photon + photo plate,
becomes exactly described by an entangled quantum state. It is
possible only under condition that before given interaction
between photon and photo plate photon has been exactly quantum
mechanically described by a superposition but not by a mixture of
the quantum states, i.e. trajectories. Thus, by exact quantum
mechanical dynamical interaction between photon and movable photo
plate there is none dynamical change of the past, precisely
quantum mechanical dynamical state of the photon before
interaction with photo plate.

In this way standard quantum mechanical formalism does not admit
exactly any delayed choice, i.e. influence at the past. Only by
incomplete sub-systemic measurements and analyses on the photon it
can be formally, i.e. effectively concluded that photon already
before interaction with photo plate has been in a mixture of the
quantum states, i.e. that photon like a classical particle
propagates always along one trajectory with corresponding
probability.

So, in the original Wheeler's delayed choice experiment, after
quantum mechanical dynamical interaction between photon and
movable photo plate there is only the following choice. For an
additional analysis either complete, super-systemic or incomplete,
sub-systemic description of the quantum super-system, photon +
movable photo plate, can be chosen. This conclusion is pointed out
more explicitly in our variant of the delayed choice experiment.

\section {References}

\begin {itemize}

\item [[1]]  J. A.Wheeler, in {\it Quantum Theory and Measurement}, eds. J. A. Wheeler, W. H. Zurek (Princeton University Press, Princeton,1984.)
\item [[2]]  J. A. Wheeler, in {\it Mathematical Foundations of Quantum Theory}, ed. A. R. Marlow (Academic Press, New York, 1978.)
\item [[3]]  V. Jacques, E. Wu, F. Grosshans, F. Treussart, P. Grangier, A. Aspect , J.-F. Roch, {\it Experimental Realization of Wheeler's delayed choice Gedanken Experiment}, quant-ph/0610241
\item [[4]]  N . Bohr, {\it Atomic Physics and Human Knowledge} (John Wiley and Sons, New York, 1958.)
\item [[5]]  N. Bohr, Phys.Rev., {\it 48}, (1935.), 696.
\item [[6]]  J. von Neumann,  {\it Mathematische Grundlagen der Quanten Mechanik} (Spiringer Verlag , Berlin , 1932.)
\item [[7]]  B. d'Espagnat,  {\it Conceptual Foundations of Quantum Mechanics} (Benjamin, New York, 1976.)
\item [[8]]  A. Aspect, P. Grangier, G. Roger, Phys.Rev.Lett., {\bf 47}, (1981.), 460.
\item [[9]]  R.P.Feynman, R. Leighton, M.Sands, {\it Feynman Lectures on Physics} (Addison-Wesley, Reading, Mass., 1963.)
\item [[10]]  R.P.Feynman, {\it The Character of Physical Law} (Cox and Wyman LTD, London, 1965.)
\item [[11]]  J. S. Bell, Physics, {\it 1}, (1964.), 195.
\item [[12]] V. Pankovi\'c, M. Predojevi\'c, M. Krmar, {\it Quantum Superposition of a Mirror and Relative Decoherence (as Spontaneous Superposition Breaking) }, quant-ph/0312015.
\item [[13]]  V. Pankovi\'c, T. H$\ddot{u}$bsch, M. Predojevi\'c,  M. Krmar, {\it From Quantum to Classical Dynamics: A Landau Phase Transition with Spontaneous Superposition Breaking}, quant-ph/0409010.
\item [[14]]  J. Bernstein, Rev.Mod.Phys., {\it 46}, ( 1974.), 7.
\item [[15]]  F. Halzen, A. Martin, {\it Quarks and Leptons : An Introductory Course in Modern Particle Physics} (John Wiley and Sons, New York, 1987.)
\item [[16]]  L. H. Ryder, {\it Quantum Field Theory} (Cambridge University Press, Cambridge, 1987.)
\item [[17]]  W. Marshall, C. Simon, R. Penrose, D. Bouwmeester, Phys.Rev.Lett., {\it 91}, (2003.), 130401.

\end {itemize}

\end {document}